\documentclass[aps,preprint]{revtex4-1}
\usepackage{graphicx}
\usepackage{amsmath}
\usepackage{makeidx}
\usepackage{amsfonts}
\usepackage{amssymb}

\begin{document}

\title{Low-temperature excitations within the Bethe approximation}

\author{I. Biazzo}
\email{indaco.biazzo@polito.it}
\affiliation{Physics Department and Center for Computational Sciences, Politecnico di Torino, Corso Duca degli Abruzzi 24, 10129 Torino, Italy}

\author{A. Ramezanpour}
\email{abolfazl.ramezanpour@polito.it}
\affiliation{Physics Department and Center for Computational Sciences, Politecnico di Torino, Corso Duca degli Abruzzi 24, 10129 Torino, Italy}

\date{\today}

\begin{abstract}
We propose the variational quantum cavity method to construct a minimal energy subspace of wave vectors that are used to obtain some upper bounds for the energy cost of the low-temperature excitations. Given a trial wave function we use the cavity method of statistical physics to estimate the Hamiltonian expectation and to find the optimal variational parameters in the subspace of wave vectors orthogonal to the lower-energy wave functions. To this end, we write the overlap between two wave functions within the Bethe approximation which allows us to replace the global orthogonality constraint with some local constraints on the variational parameters. The method is applied to the transverse Ising model and different levels of approximations are compared with the exact numerical solutions for small systems. 
\end{abstract}


\maketitle

\section{Introduction}\label{S0}
The variational problem of finding an excited quantum state can be formulated as a classical optimization problem constrained by the orthogonality conditions imposed by the lower-energy eigenstates of the Hamiltonian. This is a computationally hard task even for the ground-state problem in one-dimensional quantum systems. Exact diagonalization algorithms are very helpful but they are limited to small systems due to the exponential growth of the Hilbert space with the size of system \cite{LG-cp-1993}. For larger systems one has to resort to approximation methods, e.g., the variational quantum Monte Carlo algorithms, to study the low-energy states of the Hamiltonian \cite{N-cpl-1978,T-jpc-1979,MC-rmp-2001}. On the other hand, one can always obtain useful insights by studying some exactly solvable mean-field models \cite{BS-jstat-2012}.        

The cavity method of statistical physics provides the framework and machinery we need to deal with random optimization and constraint satisfaction problems \cite{MP-epjb-2001,MZ-pre-2002,KMRSZ-2007,MM-book-2009}.  In particular, the method provides a local message-passing algorithm to study single instances of a problem on random and finite-connectivity graphs within the hierarchy of replica symmetry breaking (RSB) approximations \cite{MZ-pre-2002,BMZ-rsa-2005,ABRZ-prl-2011}. For instance, the cavity method can be used to minimize the Bethe estimation of computationally difficult cost functions like the Hamiltonian expectation in a variational quantum problem; first the expectation is computed in the replica symmetry (RS) approximation to obtain an energy function of the variational parameters. Then, the parameters are considered as statistical variables to study their statistical properties within a higher level RS approximation, with equations that resemble the one-step RSB ones. These equations are exploited in a message-passing algorithm to optimize over the variational parameters. There are other applications of the cavity method to quantum systems that work with the density matrices or the path integral representation of the quantum partition function \cite{H-prb-2007,LP-aphys-2008,LSS-prb-2008,KRSZ-prb-2008,DM-2010}.     

In this paper we use the variational quantum cavity method \cite{R-prb-2012} to find approximate solutions for the exited-states of the transverse Ising model. The main point of this study is to write the overlap between two wave functions in the Bethe approximation, which is asymptotically exact as long as the trial wave functions can be represented by classical systems of locally tree-like interaction graphs. This allows us to replace the global orthogonality condition with some local constraints on the variational parameters and the cavity messages in the Bethe expression for the overlap. In summary, given an appropriate trial wave function, we evaluate the Hamiltonian expectation in the subspace of wave functions orthogonal to the lower-energy states of the Hamiltonian and find an estimation of the optimal variational parameters, all within the Bethe approximation and implemented by a local message-passing algorithm.

Note that, usually, for large and disordered systems we do not have the exact ground-state of the system. The best that we can do is to find a good variational wave function that represent the low-energy state of the system. Obviously, an excited state that is orthogonal to an approximate ground state does not necessarily provide an upper bound for the excited-state energy. There are, of course, other ways to find the excited states that do not rely on the orthogonality to a priori known ground state, but we found the above procedure more amenable to the cavity method that we are going to use in this paper. Nevertheless, any orthogonal set of quantum states defines a subspace of wave vectors that according to the Courant min-max theorem \cite{Courant-book} can be used to find some upper bounds for the Hamiltonian eigenvalues.

In the next section we give the definitions and write the general equations. Section \ref{S2} is devoted to the mean-field approximation, where we take product states to represent the low-energy eigenstates of the Hamiltonian. Then, in Sec. \ref{S3} we consider more complicated trial wave functions including the two-body Jastrow interactions. The numerical results are presented in Sec. \ref{S4} and the concluding remarks are given in Sec. \ref{S5}.

\section{Definitions}\label{S1}
Consider the transverse field Ising model with Hamiltonian $H=-\sum_{(ij) \in \mathcal{E}_q} J_{ij} \sigma_i^z\sigma_j^z-\sum_{i} h_i \sigma_i^x$  
with index $i=1,\dots,N$ that labels the sites in the quantum interaction graph $\mathcal{E}_q$. The $\sigma_i^{x,y,z}$ are the standard Pauli matrices. In the following we will work in the $\sigma^z$ representation with orthonormal basis $|\underline{\sigma} \rangle \equiv |\sigma_1 \sigma_2 \cdots \sigma_N \rangle$. 
Given a trial wave function $|\Psi(\underline{P})\rangle=\sum_{\underline{\sigma}} \psi(\underline{\sigma};\underline{P}) |\underline{\sigma} \rangle$ depending on a set of variational parameters $\underline{P}$, we write
the Hamiltonian expectation as
\begin{align}
\langle \Psi(\underline{P}) | H |\Psi(\underline{P})\rangle= \sum_{\underline{\sigma}} |\psi(\underline{\sigma};\underline{P})|^2 E(\underline{\sigma}), \hskip1cm E(\underline{\sigma})\equiv  \sum_{(ij) \in \mathcal{E}_q} e_{ij}(\sigma_i,\sigma_j)+ \sum_{i} e_{i}(\underline{\sigma}),
\end{align}
where $e_{ij}(\sigma_i,\sigma_j) \equiv -J_{ij} \sigma_i\sigma_j$ and 
\begin{align}
e_{i}(\underline{\sigma}) \equiv -h_i \mathrm{Re}\left\{ \sum_{\underline{\sigma}'} \frac{\psi^*(\underline{\sigma}';\underline{P})}{\psi^*(\underline{\sigma};\underline{P})} \langle \underline{\sigma}' | \sigma_i^x |\underline{\sigma} \rangle \right\}.
\end{align}
Depending on the trial wave function we obtain different expressions for the non-diagonal terms $e_{i}(\underline{\sigma})$.

We consider $\mu(\underline{\sigma};\underline{P})\equiv |\psi(\underline{\sigma};\underline{P})|^2$ as a probability measure 
in a classical system and compute the above average quantities within the Bethe approximation.
For $\psi(\underline{\sigma};\underline{P})\propto \prod_{a\in \mathcal{E}_c} \phi_a(\underline{\sigma}^{\partial a};P_a)$,
the classical measure is given by  
$\mu(\underline{\sigma};\underline{P}) \propto 
\prod_{a} |\phi_a(\underline{\sigma}^{\partial a};P_a)|^2$
with the set of classical interactions  
$\mathcal{E}_c\equiv \{\phi_a(\underline{\sigma}^{\partial a})|a=1,\dots,A\}$. 
Here $\partial a$ is the subset of variables that appear in $\phi_a$. 
Similarly we define $\partial i$ as the subset of interactions that depend on $\sigma_i$.
The cavity marginal $\mu_{i\to a}(\sigma_i)$ gives the
probability of having $\sigma_i$ in absence of interaction term $\phi_a$. 
Similarly we define $\mu_{a\to i}(\sigma_i)$ as the
probability of having $\sigma_i$ in absence of the other interactions involving $\sigma_i$. 
The equations governing these cavity marginals are:
\begin{align}
\mu_{i\to a}(\sigma_i) &\propto  \prod_{b \in \partial i \setminus a} \mu_{b\to i}(\sigma_i), \\
\mu_{a\to i}(\sigma_i) &\propto \sum_{\underline{\sigma}^{\partial a \setminus i} } |\phi_a(\underline{\sigma}^{\partial a};P_a)|^2 \prod_{j \in \partial a \setminus i} \mu_{j\to a}(\sigma_j).
\end{align}
These are the belief propagation (BP) equations \cite{KFL-inform-2001} that can be solved by iteration starting from random initial cavity marginals. In the replica-symmetric approximation we assume there is a fixed point to the BP equations describing the single Gibbs state of the system \cite{MM-book-2009}. The above cavity marginals are enough to obtain the Bethe estimation of the average energy.   

Let us denote by $|\Psi_n(\underline{P}^n)\rangle$ the $n$th trial wave function that minimizes the average energy $\langle \Psi_n(\underline{P}^n) | H |\Psi_n(\underline{P}^n)\rangle$
conditioned on the orthogonality constraints $\langle \Psi_n(\underline{P}^n) |\Psi_m(\underline{P}^m)\rangle=0$ for $m=0,\dots,n-1$. The corresponding classical systems are represented by measures $\mu_n(\underline{\sigma};\underline{P}^n)$.
In the following we are going to satisfy the orthogonality constraints within the Bethe approximation, 
\begin{eqnarray}
\langle \Psi_n(\underline{P}^n) |\Psi_m(\underline{P}^m)\rangle= \sum_{\underline{\sigma}} \psi_n^*(\underline{\sigma};\underline{P}^n) \psi_m(\underline{\sigma};\underline{P}^m) \simeq e^{-\left(\sum_i \Delta F_i + \sum_a \Delta F_a-\sum_{(ia)} \Delta F_{ia} \right)}=0, 
\end{eqnarray}
where $\Delta F_i$, $\Delta F_a$, and $\Delta F_{ia}$ are the free energy changes by adding variable node $i$, interaction node $a$, and link $(ia)$ to the complex measure $\nu_{n,m}(\underline{\sigma};\underline{P}^n,\underline{P}^m)\propto \psi_n^*(\underline{\sigma};\underline{P}^n) \psi_m(\underline{\sigma};\underline{P}^m)\propto  \prod_a  \phi_a^*(\underline{\sigma}^{\partial a};P_a^n)\phi_a(\underline{\sigma}^{\partial a};P_a^m)$. These quantities are given by \cite{MM-book-2009},
\begin{align}
e^{-\Delta F_i} &= \sum_{\sigma_i} \prod_{a \in \partial i} \nu_{a\to i}(\sigma_i), \\
e^{-\Delta F_a} &= \sum_{\underline{\sigma}^{\partial a} } \phi_a^*(\underline{\sigma}^{\partial a};P_a^n)\phi_a(\underline{\sigma}^{\partial a};P_a^m) \prod_{i \in \partial a } \nu_{i\to a}(\sigma_i)\\
e^{-\Delta F_{ia}} &= \sum_{\sigma_i} \nu_{i\to a}(\sigma_i)\nu_{a\to i}(\sigma_i),
\end{align}
where the cavity marginals $\nu_{i\to a}(\sigma_i)$ and $\nu_{a\to i}(\sigma_i)$ satisfy the BP equations for the complex measure $\nu_{n,m}(\underline{\sigma};\underline{P}^n,\underline{P}^m)$.
Thus, to have orthogonality it is enough to have $e^{-\Delta F_a}=0$ for some $a$. This defines a constraint on the parameter $P_a^n$ given $\underline{P}^m$.

In summary, to estimate the average energy and to satisfy the orthogonality constraints we need to know the BP marginals of the classical measure $\mu_n$ and $\{\nu_{n,m}| m=0,\dots,n-1\}$. In addition, we have to choose the set of constrained parameters $A_n\equiv \{a_m|m=0\dots,n-1\}$ for the orthogonality constraints $e^{-\Delta F_{a_m}}=0$.  
In this paper we will use a greedy strategy to construct $A_n$, by choosing the parameters that at least locally minimize the energy expectation.    
Finally, the problem of minimizing over the variational parameters is considered as a classical statistical physics problem 
\begin{eqnarray}
\mathcal{Z}_n=\sum_{\underline{P}^n}\sum_{\mu_n,\{\nu_{n,m}\}} \mathbb{I}_{BP}\prod_{m=0,\dots,n-1}\mathbb{I}_{n,m}e^{-\beta_{opt} \langle E(\underline{\sigma}) \rangle_{\mu_n}},
\end{eqnarray}
where for $\beta_{opt} \to \infty$ the Gibbs measure is concentrated on the optimal parameters. The indicator functions $\mathbb{I}_{BP}$ and $\mathbb{I}_{n,m}$ ensure that the messages $\mu_n,\{\nu_{n,m}\}$ satisfy the BP equations and the states $n$ and $m$ are orthogonal. Starting from $n=0$, one can find the other states one by one after solving the above optimization problem.

One may find an approximate solution to the above problem by a two-stage algorithm: Given $\{\underline{P}^m|m=0,\dots,n-1\}$ and an arbitrary set of the constrained parameters $A_n$, we run BP to find the set of marginals $\{\nu_{n,m}| m=0,\dots,n-1\}$. These are used to fix the constrained parameters in $A_n$ to satisfy the orthogonality constraints. Then we minimize the average energy $\langle E(\underline{\sigma}) \rangle_{\mu_n}$ over the remaining parameters. The above two stages are repeated to converge the algorithm.

\section{The mean-field approximation}\label{S2}
Let us start with the mean-field (MF) approximation, where the trial wave functions are represented by the product states:
\begin{align}
\psi(\underline{\sigma};\underline{B}) \propto \prod_i e^{ B_i \sigma_i },
\end{align} 
with complex parameters $B_i$. This results to $e_i(\sigma_i) = -h_i e^{-2B_i^R\sigma_i}\cos(2B_i^I\sigma_i)$ and the following classical measure 
$\mu(\underline{\sigma};\underline{B}) \propto \prod_{i} e^{ 2B_i^{R} \sigma_i}$.
By superscripts $R$ and $I$ we mean the real and imaginary part of the parameters.
Given the above measure we find 
\begin{align}
\langle e_i(\sigma_i) \rangle_{\mu} &= -h_i  \frac{\cos(2B_i^{I})}{\cosh(2B_i^{R})}, \\ 
\langle e_{ij}(\sigma_i,\sigma_j) \rangle_{\mu} &= -J_{ij} \tanh(2B_i^{R})\tanh(2B_j^{R}). 
\end{align}

We see that for non-negative $h_{i}$ the average energy is minimized by setting $B_i^{I}=0$.
Therefore, as long as we are interested in the ground state, we can set the imaginary parts to zero.
In this case, the variational problem reads 
\begin{eqnarray}
\mathcal{Z}_0=\sum_{\underline{B}} e^{-\beta_{opt} \sum_{(ij)\in \mathcal{E}_q}\langle e_{ij} \rangle_{\mu} -\beta_{opt} \sum_i \langle e_i \rangle_{\mu}},
\end{eqnarray}
The cavity marginals of the parameters in the Bethe approximation are
\begin{eqnarray}
M_{i \to j}(B_i) \propto  e^{-\beta_{opt} \langle e_i \rangle_{\mu} } \prod_{k \in \partial i \setminus j} \left( \sum_{B_k} e^{-\beta_{opt}\langle e_{ik} \rangle_{\mu}} M_{k \to i}(B_k) \right).
\end{eqnarray}
For $\beta_{opt} \to \infty$ and scaling $M_{i \to j}(B_i)=e^{-\beta_{opt} \mathcal{M}_{i \to j}(B_i)}$ we find the minsum equations \cite{KFL-inform-2001}:
\begin{align}
\mathcal{M}_{i \to j}(B_i) &= \langle e_{i} \rangle_{\mu}+ \sum_{k\in \partial i\setminus j}
\min_{ B_k } 
\left\{ \langle e_{ik} \rangle_{\mu} + \mathcal{M}_{k \to i}(B_k) \right\}.
\end{align}
The equations are solved by iteration starting from random initial messages. After each iteration we subtract a constant from the messages to have $\min_{B_i} \mathcal{M}_{i \to j}(B_i)=0$.     
Then we find the optimal parameters by minimizing the local minsum weights,
\begin{align}
B_i^0=\arg \min_{B_i} \left\{\langle e_{i} \rangle_{\mu}+ \sum_{j\in \partial i}
\min_{ B_j } 
\left\{ \langle e_{ij} \rangle_{\mu} + \mathcal{M}_{j \to i}(B_j) \right\} \right\}.
\end{align} 
 
Note that in the MF approximation the orthogonality condition reads
\begin{align}
\langle \Psi_n(\underline{B}^n) |\Psi_m(\underline{B}^m)\rangle \propto \prod_i \cosh(B_i^{n*}+B_i^m)=0,
\end{align} 
thus, it is enough to have $\cosh(B_i^{n*}+B_i^{m})=0$ for some $i$. This means $B_i^{nR}=-B_i^{mR}$ and $B_i^{nI}-B_i^{mI}=\pi/2$.
Therefore, the $n$th excited state can be obtained by the following set of constraints:
\begin{align}
\mathbb{I}_{n,m}=
\left\{
                   \begin{array}{ll}
                     B_{i_m}^{nR}+B_{i_m}^{mR}=0, & \hbox{} \\
                     B_{i_m}^{nI}-B_{i_m}^{mI}=\frac{\pi}{2}, & \hbox{}
                   \end{array}
                 \right. 
\end{align} 
for $m=0,\dots,n-1$. The index $i_m$ is chosen to minimize the local energy $\langle e_{i_m}(\sigma_i) \rangle_{\mu_n}$. 
Then, one can use the same minsum equations as above to minimize over the remaining parameters. 
In this way we can find at most $N$ orthogonal product states of minimum energies $E_n$. 

Given the states $|\Psi_m(\underline{B}^m)\rangle$ for $m=0,\dots,n$, we can easily compute the Hamiltonian matrix elements $H_{mm'}\equiv \langle \Psi_m(\underline{B}^m)|H|\Psi_{m'}(\underline{B}^{m'})\rangle$, which for arbitrary parameters read 
\begin{multline}
H_{mm'}=\prod_i \left( \frac{\cosh(B_i^{m*}+B_i^{m'})}{(\cosh(2B_i^{mR})\cosh(2B_i^{m'R}))^{1/2}} \right) \\ \times  \left\{-\sum_{(ij)\in \mathcal{E}_q} J_{ij} \tanh(B_i^{m*}+B_i^{m'})\tanh(B_j^{m*}+B_j^{m'})-\sum_i h_i \frac{\cosh(B_i^{m*}-B_i^{m'})}{\cosh(B_i^{m*}+B_i^{m'})}\right\}.
\end{multline} 
One can diagonalize the Hamiltonian in the subspace spanned by the above states to obtain the eigenvalues $\lambda_m$. Then, using the 
min-max principle, we know that $\lambda_n$ is an upper bound for the $n$th eigenvalue of the Hamiltonian. Indeed, we found that in this case $\lambda_n \simeq E_n$ for large $N$ as the off-diagonal matrix elements $H_{mm'}$ decay exponentially with the size of system.    
 
Notice that instead of imposing the orthogonality exactly we could ask for an exponentially small overlap $\langle \Psi_n(\underline{B}^{n}) |\Psi_m(\underline{B}^m)\rangle < \epsilon^N$ by demanding
\begin{align}
\frac{\cosh(B_i^{n*}+B_i^m)}{(\cosh(2B_i^{nR})\cosh(2B_i^{mR}))^{1/2}}<\epsilon \le 1,
\end{align} 
for all $i$, which can easily be imposed in the above minsum equations. 
Indeed to apply the min-max theorem we do not need a set of orthogonal states \cite{Courant-book}; according to the theorem, the $n$th eigenvalue is given by $\min_{S_{n+1}} \max_{|\psi\rangle \in S_{n+1}:\langle \psi|\psi\rangle=1} \langle \psi|H|\psi\rangle$ where $S_{n+1}$ is any subspace of dimension $n+1$.

\section{Beyond the mean-field approximation}\label{S3}
We can do better than the MF approximation by adding the local two-body or Jastrow interactions \cite{J-pr-1955} to the trial wave functions:
\begin{align}
\psi(\underline{\sigma};\underline{P}) &\propto \prod_i \phi_i(\sigma_i)\prod_{(ij)\in \mathcal{E}_c}
\phi_{ij}(\sigma_i,\sigma_j),
\end{align}
with 
\begin{align}
\phi_i(\sigma_i) \equiv  e^{ B_i \sigma_i},\hskip1cm
\phi_{ij}(\sigma_i,\sigma_j) \equiv  e^{K_{ij} \sigma_i\sigma_j}.
\end{align}
For simplicity we are going to assume $\mathcal{E}_c=\mathcal{E}_q$.
As a result, we obtain 
\begin{equation}
e_{i}(\sigma_i,\underline{\sigma}^{\partial i}) =  -h_{i} 
e^{-2B_i^R\sigma_i-\sum_{j \in \partial i} 2K_{ij}^R\sigma_i\sigma_j}\cos(2B_i^I\sigma_i+\sum_{j \in \partial i} 2K_{ij}^I\sigma_i\sigma_j).
\end{equation}
The average energy is computed with respect to the following classical measure \\ $\mu(\underline{\sigma}) \propto \prod_i |\phi_i(\sigma_i)|^2\prod_{(ij)\in \mathcal{E}_c}
|\phi_{ij}(\sigma_i,\sigma_j)|^2$.
To estimate the average energies $\langle e_{ij}(\sigma_i,\sigma_j) \rangle_{\mu}$ and  $\langle e_{i}(\sigma_i,\underline{\sigma}^{\partial i}) \rangle_{\mu}$ we need the following local marginals
\begin{align}
\mu_{ij}(\sigma_i,\sigma_j)= \frac{1}{Z_{ij}}   |\phi_{ij}(\sigma_i,\sigma_j)|^2 \mu_{i \to j}(\sigma_i)\mu_{j \to i}(\sigma_j),\\
\mu_{i,\partial i}(\sigma_i,\underline{\sigma}^{\partial i})=\frac{1}{Z_{i,\partial i}}  |\phi_i(\sigma_i)|^2 \prod_{j \in \partial i}|\phi_{ij}(\sigma_i,\sigma_j)|^2 \mu_{j \to i}(\sigma_j),
\end{align}
given in terms of the cavity marginals   
\begin{align}
\mu_{i \to j}(\sigma_i) \propto  |\phi_i(\sigma_i)|^2\prod_{k \in \partial i \setminus j}\left( \sum_{\sigma_k} |\phi_{ik}(\sigma_i,\sigma_k)|^2 \mu_{k \to i}(\sigma_k) \right).
\end{align}

For $B_i=0$ we can simplify the equations by taking the symmetric (or paramagnetic) solution of the BP equations. This is of course exact for a tree classical interaction graph $\mathcal{E}_c$.  Then, the average local energies are given by
\begin{align}
\langle e_i(\sigma_i,\underline{\sigma}^{\partial i}) \rangle_{\mu} &= -h_i \prod_{j\in \partial i} \left( \frac{\cos(2K_{ij}^I)}{\cosh(2K_{ij}^R)} \right), \\ 
\langle e_{ij}(\sigma_i,\sigma_j) \rangle_{\mu} &= -J_{ij} \tanh(2K_{ij}^R). 
\end{align}
The resulting minsum equations are
\begin{align}
\mathcal{M}_{i \to j}(K_{ij}) &= 
\min_{\{K_{ik}| k \in \partial i \setminus j\}} 
\left\{ \langle e_i \rangle_{\mu} + \sum_{k \in \partial i \setminus j} \left(\langle e_{ik} \rangle_{\mu}+ \mathcal{M}_{k \to i}(K_{ik}) \right) \right\},
\end{align} 
and the optimal couplings are estimated by 
\begin{eqnarray}
K_{ij}^0= \arg \min_{K_{ij}}\left\{ \langle e_{ij} \rangle_{\mu} + \mathcal{M}_{i \to j}(K_{ij})+  \mathcal{M}_{j \to i}(K_{ij})\right\}.  
\end{eqnarray}

The orthogonality condition for two symmetric states reads
\begin{align}
\langle \Psi_n(\underline{K}^n) |\Psi_m(\underline{K}^m)\rangle \propto \sum_{\underline{\sigma}}  e^{\sum_{(ij)\in \mathcal{E}_c} (K_{ij}^{n*}+K_{ij}^m) \sigma_i\sigma_j}\propto \prod_{(ij)\in \mathcal{E}_c} \cosh(K_{ij}^{n*}+K_{ij}^{m})=0,
\end{align} 
using the symmetric solution of the BP equations.  
Thus, to have orthogonality we need $\cosh(K_{ij}^{n*}+K_{ij}^{m})=0$ for some link $(ij)$. That is, for the $n$th excited state we have the following set of constraints:
\begin{align}
\mathbb{I}_{n,m}=
\left\{
                   \begin{array}{ll}
                     K_{i_mj_m}^{nR}+K_{i_mj_m}^{mR}=0, & \hbox{} \\
                     K_{i_mj_m}^{nI}-K_{i_mj_m}^{mI}=\frac{\pi}{2}, & \hbox{}
                   \end{array}
                 \right. 
\end{align}
for $m=0,\dots,n-1$. The link $(i_mj_m)$ is chosen to minimize the local energy $\langle e_{i_mj_m} \rangle_{\mu_n}$. Then, one can use the same minsum equations as above to minimize over the remaining parameters. The number of orthogonal states that we can find in this way is limited by the number of the coupling parameters in the classical system which for a tree structure is $N-1$. The Hamiltonian matrix elements are given by
\begin{multline}
H_{mm'}=\prod_{(ij)\in \mathcal{E}_q} \left( \frac{\cosh(K_{ij}^{m*}+K_{ij}^{m'})}{(\cosh(2K_{ij}^{mR})\cosh(2K_{ij}^{m'R}))^{1/2}} \right) \\ \times  \left\{-\sum_{(ij)\in \mathcal{E}_q} J_{ij} \tanh(K_{ij}^{m*}+K_{ij}^{m'})-\sum_i h_i \prod_{j\in \partial i}\left(\frac{\cosh(K_{ij}^{m*}-K_{ij}^{m'})}{\cosh(K_{ij}^{m*}+K_{ij}^{m'})}\right)\right\}.
\end{multline}

Similarly, we can make more general orthogonal states $|\Psi_n(\underline{B}^n,\underline{K}^n)\rangle$ and $|\Psi_n(\underline{B}^m,\underline{K}^m)\rangle$ by choosing $B_{i}^n$ such that for the complex measure $\nu_{nm}\propto e^{ \sum_i (B_i^{n*}+B_i^{m})\sigma_i + \sum_{(ij)\in \mathcal{E}_c}(K_{ij}^{n*}+K_{ij}^{m})\sigma_i\sigma_j}$ we have $e^{\Delta F_i}=0$ for some $i$. This gives
\begin{align}
e^{2(B_{i}^{n*}+B_{i}^{m})}=-\prod_{j\in \partial i}\left( \frac{\sum_{\sigma_j}e^{-(K_{ij}^{n*}+K_{ij}^{m})\sigma_j} \nu_{j\to i}(\sigma_j)}{\sum_{\sigma_j}e^{(K_{ij}^{n*}+K_{ij}^{m})\sigma_j} \nu_{j\to i}(\sigma_j)} \right).
\end{align}
The node $i$ can be chosen in a greedy way to minimize $\langle e_{i}(\sigma_i,\underline{\sigma}^{\partial i}) \rangle_{\mu_n}$ in the mean-field approximation.
Here it is more difficult to minimize the average energy $\langle E(\sigma)\rangle_{\mu_n}$, which depends not only on the variational parameters but also on the BP cavity marginals $\mu_{i\to j}(\sigma_i)$. More precisely, the minsum equations read
\begin{align}
\mathcal{M}_{i \to j}(K_{ij},\mu_{ij}) &= 
\min_{B_i,\{K_{ik},\mu_{ik}| k \in \partial i \setminus j\}: \mathbb{I}_{BP}^{(i)}} 
\left\{ \langle e_i \rangle_{\mu} + \sum_{k \in \partial i \setminus j} \left(\langle e_{ik} \rangle_{\mu}+ \mathcal{M}_{k \to i}(K_{ik},\mu_{ik}) \right) \right\},
\end{align} 
where $\mu_{ij}\equiv (\mu_{i\to j},\mu_{j\to i})$. Note that the minimum in the right hand side is conditioned on satisfying the local BP equations. The reader can find more details in Ref. \cite{R-prb-2012}.

\begin{figure}
\includegraphics[width=14cm,height=5cm]{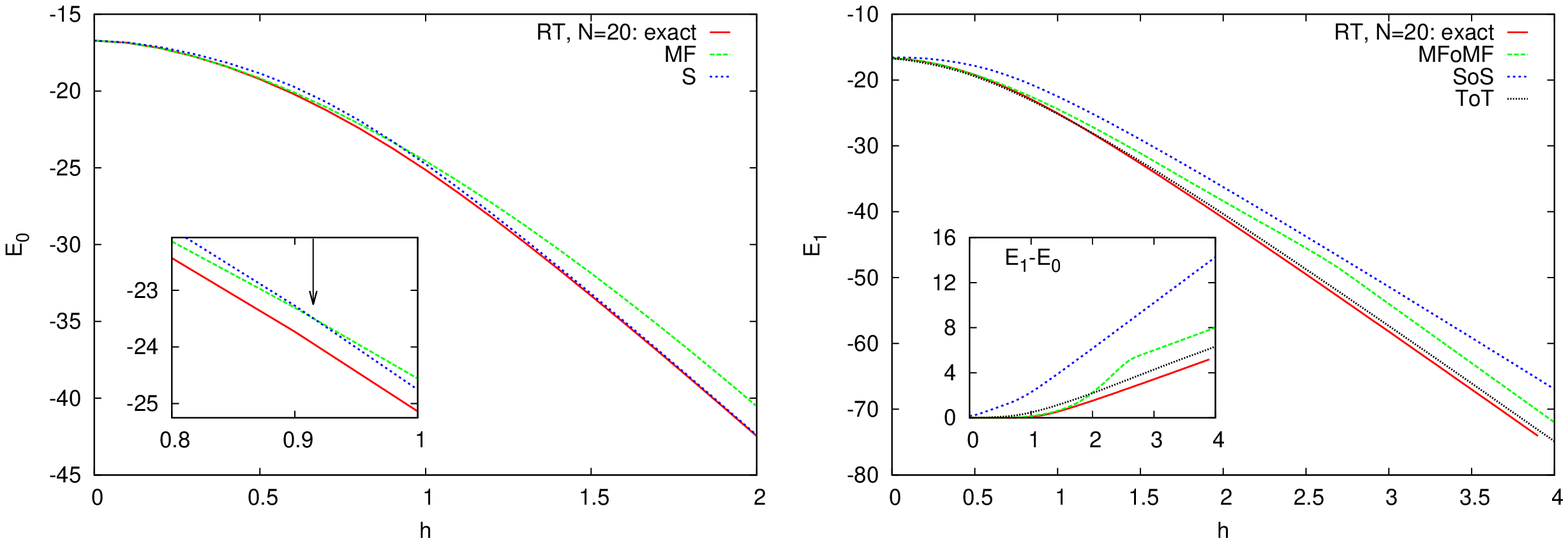}
\caption{The ground-state and excited-state energies ($E_0$ and $E_1$) for the transverse Ising model on a random tree (RT) of size $N=20$ with random Gaussian couplings $J_{ij}$ of mean zero and variance one in a uniform transverse field $h_i=h$. The exact results are compared with the upper bounds that are obtained by a minimal energy subspace spanned by product states (MFoMF), symmetric states (SoS), and tree states (ToT).}\label{f1}
\end{figure}

\section{Numerical results}\label{S4}
Let us start with a small system to compare the above approximations with the exact results for the ground and excited states. 
We take a random tree with random Gaussian couplings $J_{ij}$ of mean zero and variance one in uniform transverse fields $h_i=h$.
This system displays a phase transition from the ordered phase for $h<h_c$ with a nonzero Edwards-Anderson order parameter $q\equiv \sum_i \langle \sigma_i^z \rangle^2/N$ to a disordered phase for $h>h_c$.

\begin{figure}
\includegraphics[width=14cm,height=5cm]{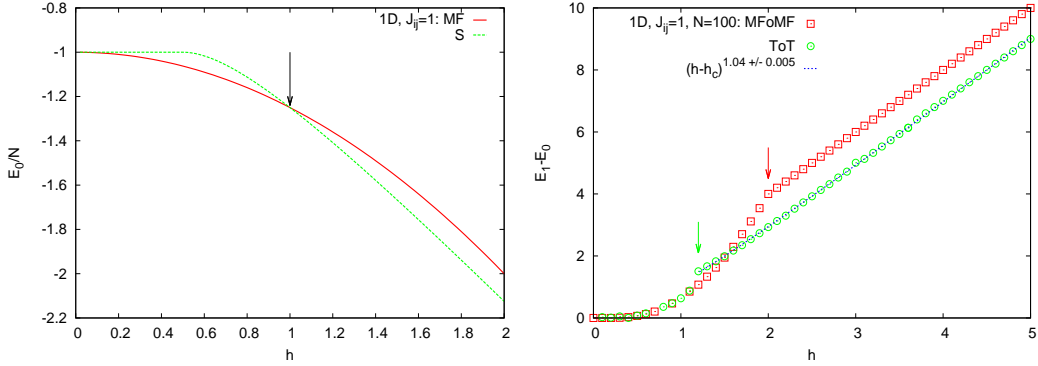}
\caption{ Left panel: The ground-state energy density ($E_0/N$) for the one-dimensional ($1$D) transverse Ising model with uniform and ferromagnetic couplings ($J_{ij}=1$) obtained by the product states (MF) and the symmetric states (S) in the thermodynamic limit.  Right panel: The energy gap $E_1-E_0$ obtained by the product states (MFoMF) and tree states (TOT) for the same model with $N=100$ spins. In both the cases the gap is non-analytic at the corresponding phase transition point, but it is non-vanishing close to the transition due to the local nature of the orthogonality constraints.}\label{f2}
\end{figure}
\begin{figure}
\includegraphics[width=14cm,height=5cm]{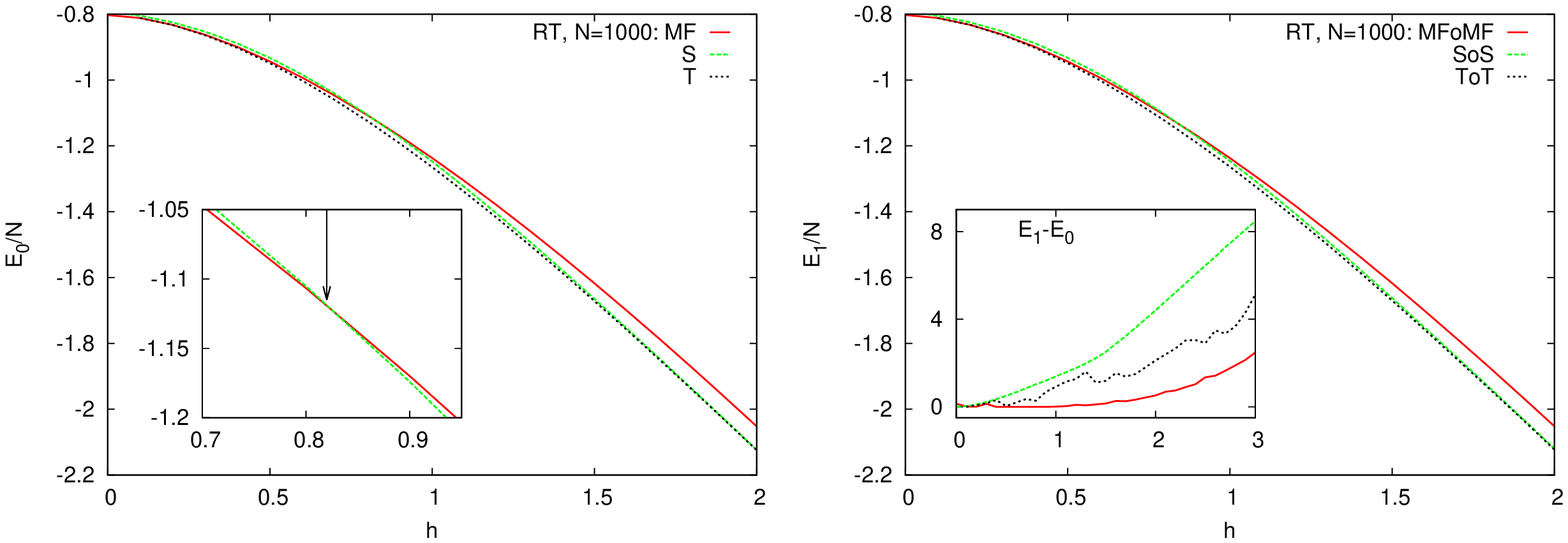}
\caption{The ground-state and excited-state energy densities ($E_0/N$ and $E_1/N$) for the transverse Ising model on a random tree (RT) of size $N=1000$ with random Gaussian couplings $J_{ij}$ of mean zero and variance one in a uniform transverse field $h_i=h$. We compare the upper bounds that are obtained by a minimal energy subspace spanned by product states (MFoMF), symmetric states (SoS), and tree states (ToT).}\label{f3}
\end{figure}

As Fig. \ref{f1} shows, we obtain better ground-state energies with the product and symmetric states in the ordered and disordered phases, respectively. While the product states allow for a nonzero magnetization, the symmetric states have by definition zero magnetization and therefore more appropriate to represent the disordered ground state.
Indeed, an estimate of the transition point can be obtained by comparing the ground-state energies that are computed by the product and symmetric states. Figure \ref{f2} displays these energies for the one-dimensional transverse Ising model with ferromagnetic couplings. Nevertheless, for small system sizes we always obtain better energies for the excited state by a minimal energy subspace of the product states.  In Fig. \ref{f1} we also display the results obtained by more general trial wave functions having a tree structure defined by $\mathcal{E}_c=\mathcal{E}_q$. As expected, we obtain much better upper bounds by introducing both the variational parameters $B_i$ and $K_{ij}$.    
Figure \ref{f3} shows the results for a larger number of spins. Here, in the disordered phase we find better upper bounds for the excited-state energy by the symmetric trial wave function. This is due to the presence of very small couplings that reduce the cost of the orthogonality constraint.

\begin{figure}
\includegraphics[width=14cm,height=5cm]{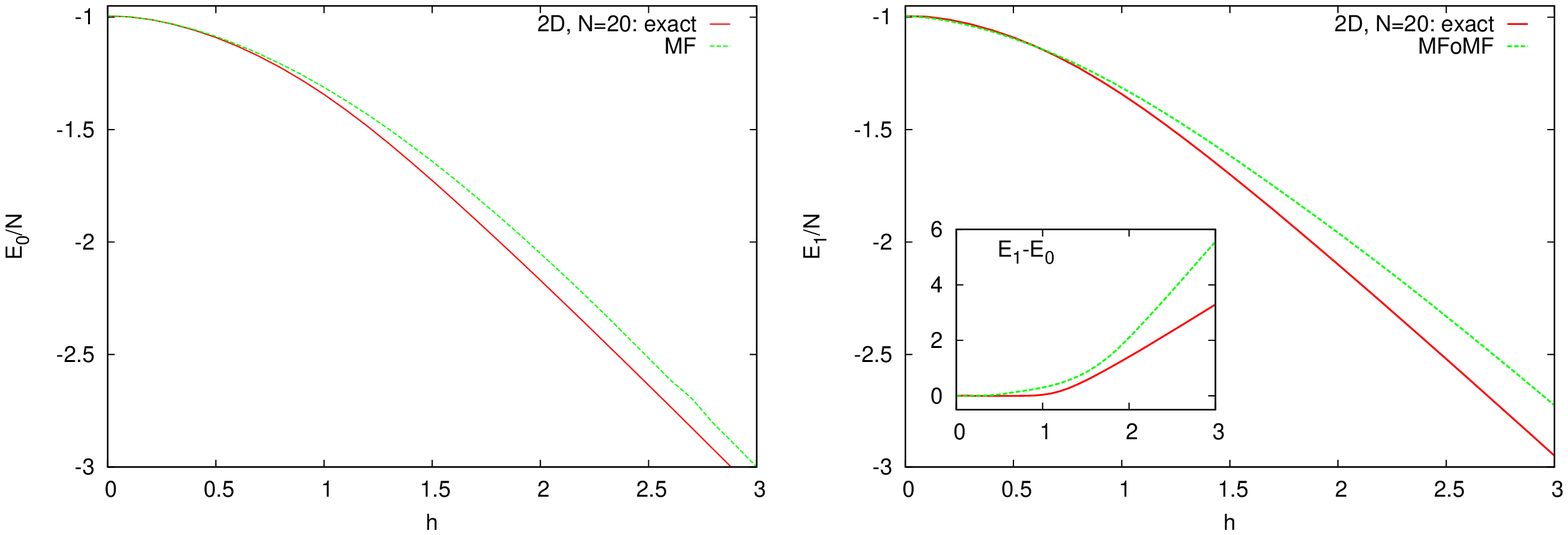}
\caption{The ground-state and excited-state energy densities ($E_0/N$ and $E_1/N$) for the transverse Ising model on a two-dimensional ($2$D) square lattice of size $N=4\times 5$ with random Gaussian couplings $J_{ij}$ of mean zero and variance one in a uniform transverse field $h_i=h$. The exact results are compared with the upper bounds that are obtained by a minimal energy subspace spanned by product states (MFoMF).}\label{f4}
\end{figure}

Finally, we present the MF results for the transverse Ising model in a two-dimensional square lattice. In Fig. \ref{f4} we compare the upper bounds for the ground and excited states with the exact ones in a small lattice.

\section{Conclusion}\label{S5}
We generalized the variational quantum cavity method to study low-temperature excitations of 
quantum systems within the Bethe approximation.  We constructed orthogonal sets of minimal energy quantum states, where the Hamiltonian matrix elements can be computed exactly to obtain some upper bounds for the Hamiltonian eigenvalues. For more general trial wave functions we have only an approximate estimation of the Hamiltonian matrix elements but the estimation is expected to be asymptotically exact as long as the trial wave functions are represented by locally tree-like classical interaction graphs. And finally, the method can also be extended to include some appropriate global interactions in the trial wave functions that are essential to deal with the fermion sign problem \cite{RZ-prb-2012}.

\acknowledgments
We would like to thank G. Semerjian and M. Zamparo for helpful discussions. A.R. acknowledges support from ERC Grant No. OPTINF  267915.

\end{document}